\documentclass[manuscript]{aastex}
\usepackage{lscape}
\usepackage{rotating}
\usepackage{color}
\shorttitle{Dome A meteorological data for astronomy} 
\shortauthors{Hu et al.}

\title{Meteorological data for the astronomical site at Dome A, Antarctica}

\def\wisk#1{\relax\ifmmode{#1}\else{$#1$}\fi} 
\def\arcseconds{\wisk{\,^{\prime\prime}}}
\def\metre{\wisk{\,\rm m}}
\def\km{\wisk{\,\rm km}}
\def\mps{\wisk{\,\rm m/s}}
\def\etal{et al.\@}

\begin{document}

\author{Yi Hu\altaffilmark{1}, Zhaohui Shang\altaffilmark{2,1}, Michael C.
B. Ashley\altaffilmark{3}, Colin S. Bonner\altaffilmark{3}, \\
Keliang Hu\altaffilmark{1}, Qiang Liu\altaffilmark{1}, Yuansheng
Li\altaffilmark{4}, Bin Ma\altaffilmark{1},  Lifan
Wang\altaffilmark{5,6},
and Haikun Wen\altaffilmark{7} }

\email{zshang@gmail.com}

\altaffiltext{1}{National Astronomical Observatories, Chinese Academy of Sciences, Beijing 100012, P. R. China}
\altaffiltext{2}{Tianjin Astrophysics Center, Tianjin Normal University, Tianjin 300387, P. R. China}
\altaffiltext{3}{School of Physics, University of New South Wales, NSW 2052, Australia}
\altaffiltext{4}{Polar Research Institute of China, Pudong, Shanghai 200136, P. R. China}
\altaffiltext{5}{Purple Mountain Observatory, Nanjing 210008, P. R. China}
\altaffiltext{6}{Physics Department, Texas A\&M University, College Station, TX 77843}
\altaffiltext{7}{Nanjing Institute of Astronomical Optics and Technology, Nanjing 210042, P. R. China}

\begin{abstract}

We present an analysis of the meteorological data collected at Dome A,
Antarctica by the Kunlun Automated Weather Station, including
temperatures and wind speeds at eight elevations above the snow
surface between 0\metre\ and 14.5\metre.  The average temperatures at
2\metre\ and 14.5\metre\ are $-54^{\circ}$C and $-46^{\circ}$C,
respectively.  We find that a strong temperature inversion existed at
all heights for more than 70\% of the time, and the temperature
inversion typically lasts longer than 25 hours, indicating an
extremely stable atmosphere.  The temperature gradient is larger at
lower elevations than higher elevations.  The average wind speed was
1.5\mps\ at 4\metre\ elevation.  We find that the temperature
inversion is stronger when the wind speed is lower and the temperature
gradient decreases sharply at a specific wind speed for each
elevation.  The strong temperature inversion and low wind speed
results in a shallow and stable boundary layer with weak atmospheric
turbulence above it, suggesting that Dome A should be an excellent
site for astronomical observations. All the data from the weather
station are available for download.

\end{abstract}


\section{Introduction}

Dome A---the highest location on the Antarctic Plateau at an elevation
of 4,089m---has long been considered to be an excellent site for
astronomical observations over a broad range of the electromagnetic
spectrum (e.g., Gillingham 1991, Marks 2002, Burton 2010, and
references therein). Dome A's advantages include extremely low
temperatures, an extremely dry atmosphere (Sims \etal\ 2012), low wind
speed, and long periods of continuous cloud-free nighttime.

Other sites in Antarctica, such as the South Pole itself, Dome C, and
Dome Fuji, are also promising, and have been extensively studied over
many years with a variety of instruments. Astronomers are particularly
interested in the layer within tens or hundreds of meters of the
surface over the Antarctic plateau since a number of studies (e.g.,
Marks \etal\ 1999, Lawrence \etal\ 2004) have shown that the
transition to the free atmosphere occurs within this height range, as
opposed to 1--2\km\ from typical observatory sites at temperate
latitudes.

At the South Pole, Marks \etal\ (1996) used microthermal sensors on a
27\metre\ mast to measure the turbulence in the lowest part of the boundary
layer at South Pole. Marks \etal\ (1999) extended the altitude
coverage by using balloon-borne microthermals to estimate a
free-atmosphere seeing component of 0.37\arcseconds\  above a boundary layer
with an average thickness of $\sim$220\metre. Travouillon \etal\ (2003)
confirmed these results using a sonic radar able to reach heights of
890\metre\ above the surface, and derived a ground-based average seeing of
1.73\arcseconds, dropping to 0.37\arcseconds\ above 300\metre. 

At Dome C, Lawrence \etal\ (2004) used a MASS-DIMM instrument and a
SODAR to measure a median free atmospheric seeing of
0.23\arcseconds\ above a boundary layer that was less than
30\metre\ high. Aristidi \etal\ (2005) analyzed two decades of
temperature and wind speed data from an automated weather station, and
four summers of measurements with balloon-borne sondes. Trinquet
\etal\ (2008) used balloon-borne measurements to find that the median
value of the boundary layer height at Dome C is 33\metre.

At Dome Fuji, astronomical seeing at times below 0.2\arcseconds\ has
been reported from a DIMM instrument 11\metre\ above the snow surface
(Okita \etal\ 2013).

Dome A was first visited by a Chinese Antarctica Research Expedition
(CHINARE) team in 2005, so opportunities for studying the site
conditions have not existed for as long as for the other sites in
Antarctica.  Following the repeated visits in recent years by CHINARE
teams, and the construction of the Kunlun Station at Dome A in 2009,
several site-testing experiments have been installed (e.g., Yang
\etal\ 2009) and encouraging results have been reported (e.g., Yang
\etal\ 2010).  For example, using data from a sonic radar, Bonner
\etal\ (2010) found that the median boundary layer thickness at Dome A
between February and August in 2009 was only 13.9\metre, considerably
lower than the median value of 33\metre\ at Dome C (Trinquet
\etal\ 2008) and 200--300\metre\ at South Pole (Travouillon
\etal\ 2003).

Missing up until now has been a detailed analysis of the meteorological
conditions at Dome A from mast-based instruments. Such measurements
can give direct information on the atmospheric turbulence and wind
speed in the boundary layer, and are relevant to the operating
environment of future telescopes, and the performance of adaptive
optics systems (Aristidi \etal\ 2005).  In this article, we present
the first results from analyzing the weather data collected
from the Kunlun Automated Weather Station (KL-AWS) at Dome A. Data
acquisition and reduction are described in \S2.  A statistical
analysis of temperature and wind speed data are presented in \S3.
Finally, a discussion is given in \S4.

\section{Data acquisition and reduction}

\begin{figure*}
\includegraphics[width=0.4\columnwidth]{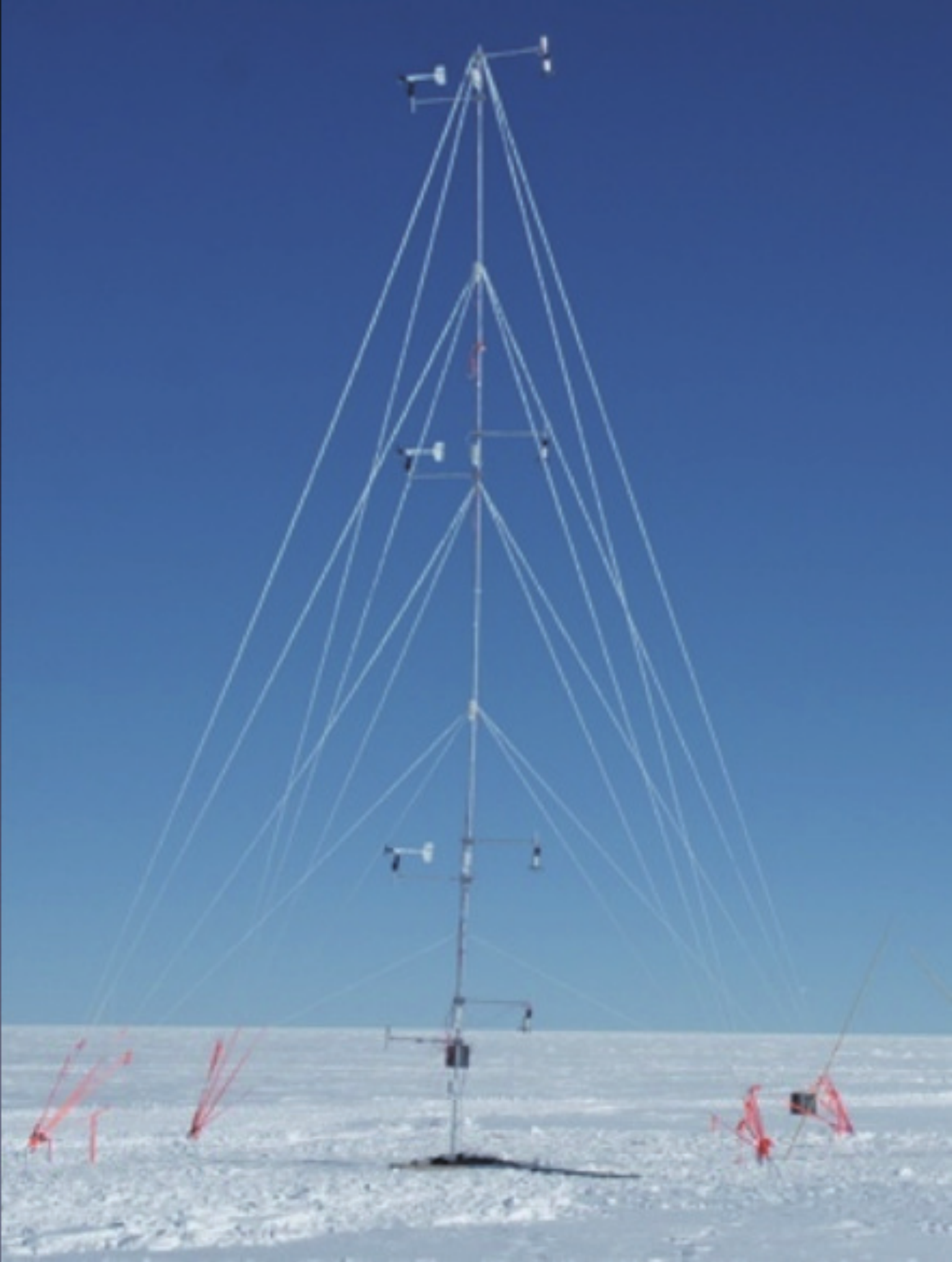}
\caption{The Kunlun Automated Weather Station (KL-AWS) installed at Dome A in January 2011. The propeller anemometers at 2\metre, 4\metre, 9\metre\ and 14.5\metre\ are clearly visible.}
\label{fig:fig1}
\end{figure*}

KL-AWS was installed in 2011 January by the 27th CHINARE team (see
Fig.\@ \ref{fig:fig1}) at Dome A at the location 80$^\circ$
25$^\prime$ 03\arcseconds~S, 77$^\circ$ 05$^\prime$
32\arcseconds~E. KL-AWS consists of a 15\metre\ mast equipped with
nine Young 4-wire 41342 temperature sensors at heights of $-$1\metre,
0\metre, 1\metre, 2\metre, 4\metre, 6\metre, 9.5\metre, 12\metre\ and
14.5\metre, four Young Wind Monitor-AQ model 05305V propeller
anemometers at heights of 2\metre, 4\metre, 9\metre\ and
14.5\metre\ to measure both wind speed and direction, and one Young
61302V barometer at 1\metre.  The manufacturer's specified absolute
measurement accuracy of the instruments is: temperature
$0.32^{\circ}$C, wind speed 0.2\mps, and wind direction $0.5^{\circ}$
plus installation offset. Originally, the mast was planned to have
four Young 85000 2D sonic anemometers, but an electrical fault
prevented their operation.  The anemometer at 2\metre\ did not acquire
data for unknown reasons.  KL-AWS was powered by PLATO, an automated
observatory (Lawrence \etal\ 2008), and it was connected to the
supervisor computers in PLATO's instrument module. KL-AWS was placed
30\metre\ south of PLATO to minimize any influence from nearby
structures. KL-AWS acquired data from all sensors once every 36
seconds and stored the data on the hard disks of the supervisor
computer.  The data were sent back in real-time every 20 minutes
through the Iridium satellite network in order to assist with
monitoring and operating the other experiments on site.

KL-AWS began operation on 2011 January 17 and ran continuously for a
year apart from a 20 day gap beginning on 2011 August 3 due to a
computer script inadvertently stopping. The sampling rate was 
$1/36$Hz up until 2011 August 3, and $1/60$Hz from 2011 August 23.
Since then, all the temperature sensors and the barometer were 
still working but 
the anemometer at 14.5\metre\ was never recovered, and other 
two anemometers at 4\metre\ and 9\metre\ were recovered 
on October 13. On 2012 January 18, KL-AWS ceased operation, possibly as a
result of damage to the power and data cables connecting it to PLATO.

\begin{figure*}
\includegraphics[width=\columnwidth]{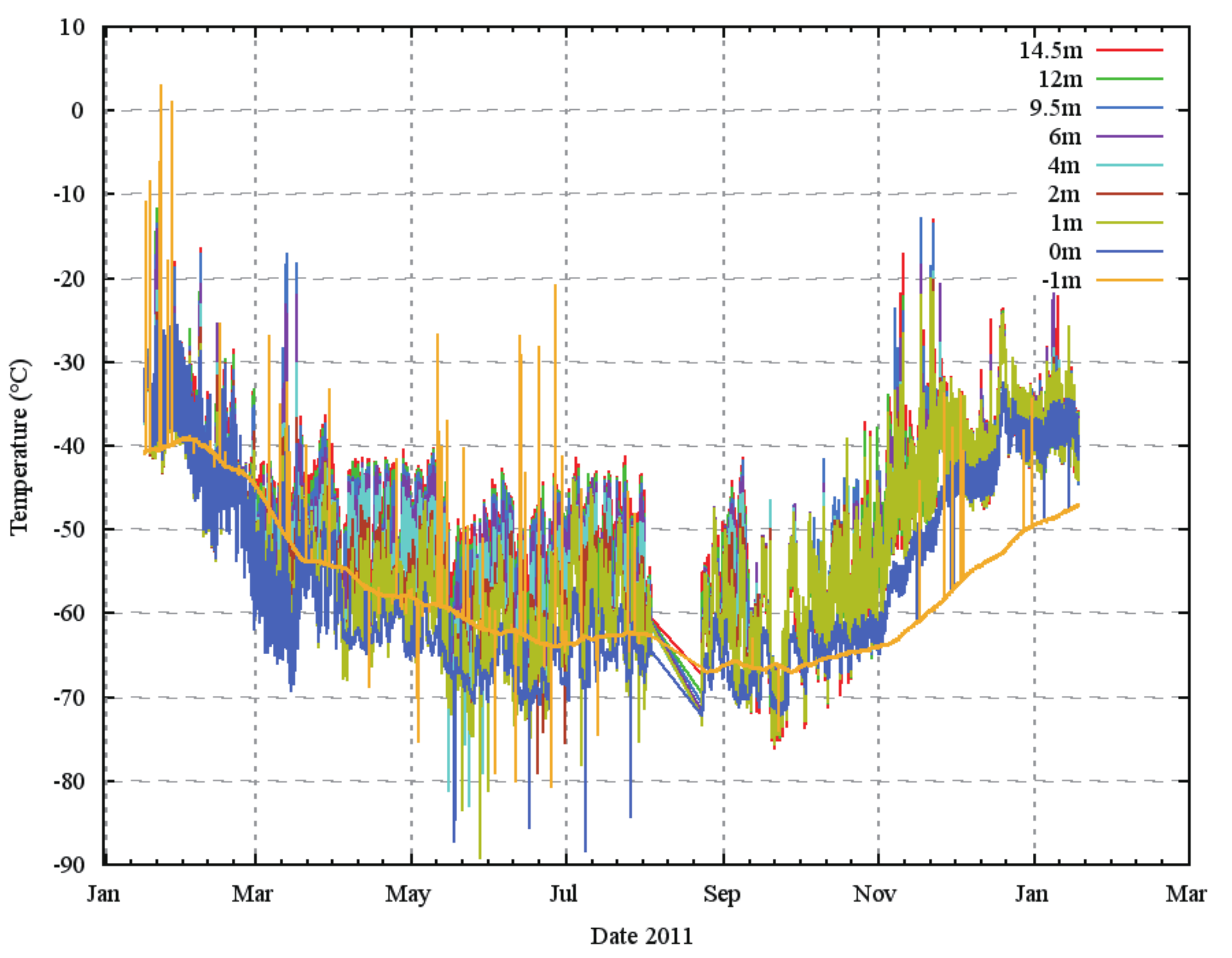}
\caption{Temperatures at nine heights, from 1\metre\ under the snow surface to 14.5\metre\ above the snow. The smoothed data (see \S2) are shown in this figure. The raw data are available for download (see \S4).}
\label{fig:fig2}
\end{figure*}

Fig.\@ \ref{fig:fig2} shows all the temperature data collected by
KL-AWS.

When the CHINARE expedition revisited Dome A at the beginning of 2012 it
was found that the top section of the weather tower had bent over and
almost touched the ground. From looking at the data we determined that
the tower likely failed on 2011 September 16 (see Fig.\@
\ref{fig:fig3}). Before that date there were often large temperature
differences between the sensors, but after September 16 the higher
sensors showed no differences.  Unless otherwise stated, this paper
restricts itself to the continuous period of data up until 2011 August
3.

To reduce the short-term fluctuations in the data, we have smoothed
the temperature data by a boxcar of 10 adjacent data points.  Fig.\@
\ref{fig:fig4} shows an example of the results where the short-term
fluctuations are removed while the temperature trends remains.  All
the following analyses are based on the smoothed temperature data.
 
\section{Results}
\subsection{Temperature and Temperature Difference\label{sec:t}}

\begin{figure*}
\includegraphics[width=\columnwidth]{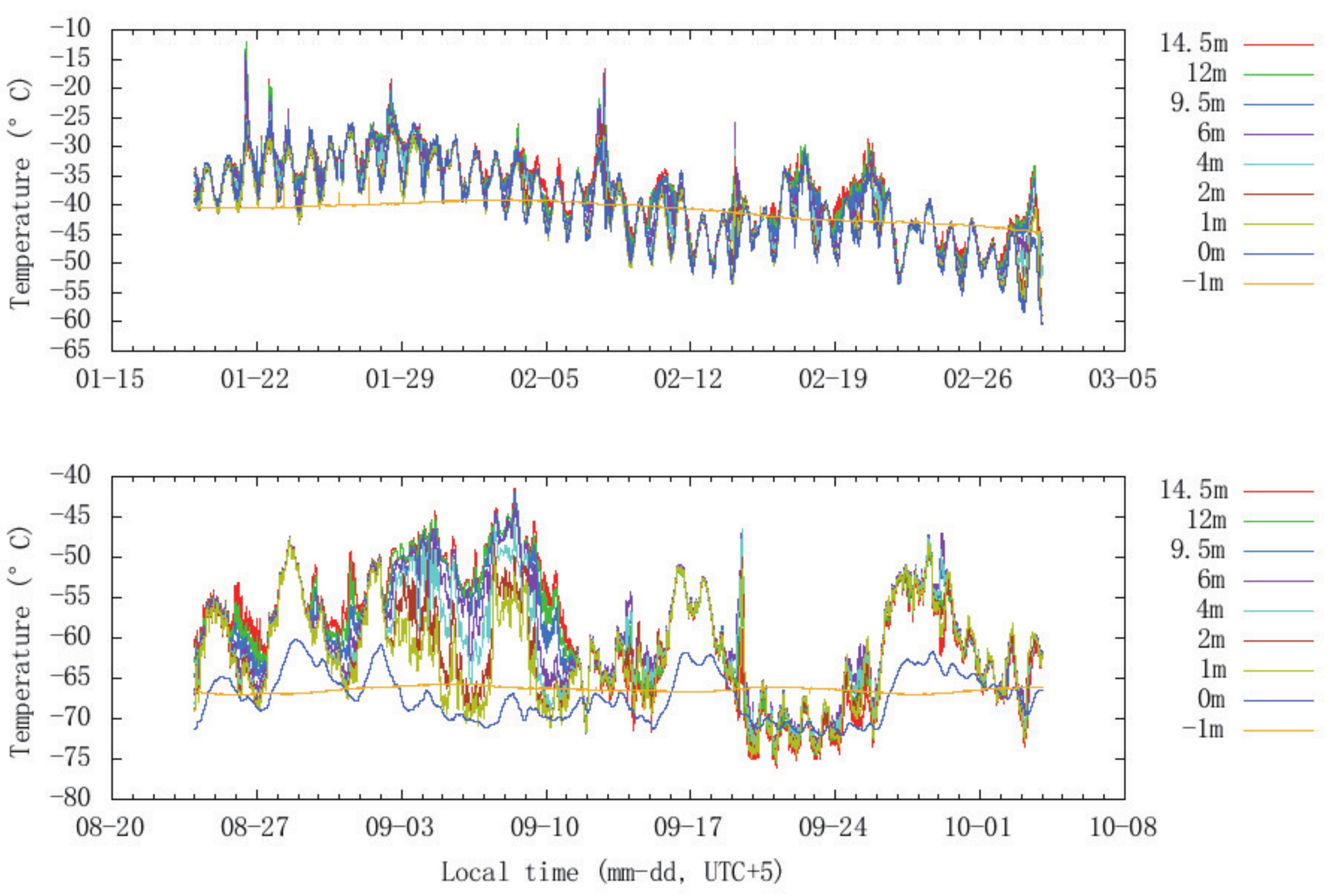}
\caption{Temperatures recorded during the first 40 days of KL-AWS operation (top panel) in 2011 and 40 days following a restart of KL-AWS on 2011 August 23 (bottom panel). The changes in the temperature distribution from 2011 September 16 are believed to have been caused by the collapse of the upper part of the mast. This figure shows smoothed data (see \S2).}
\label{fig:fig3}
\end{figure*}

\begin{figure*}
\includegraphics[width=0.5\columnwidth]{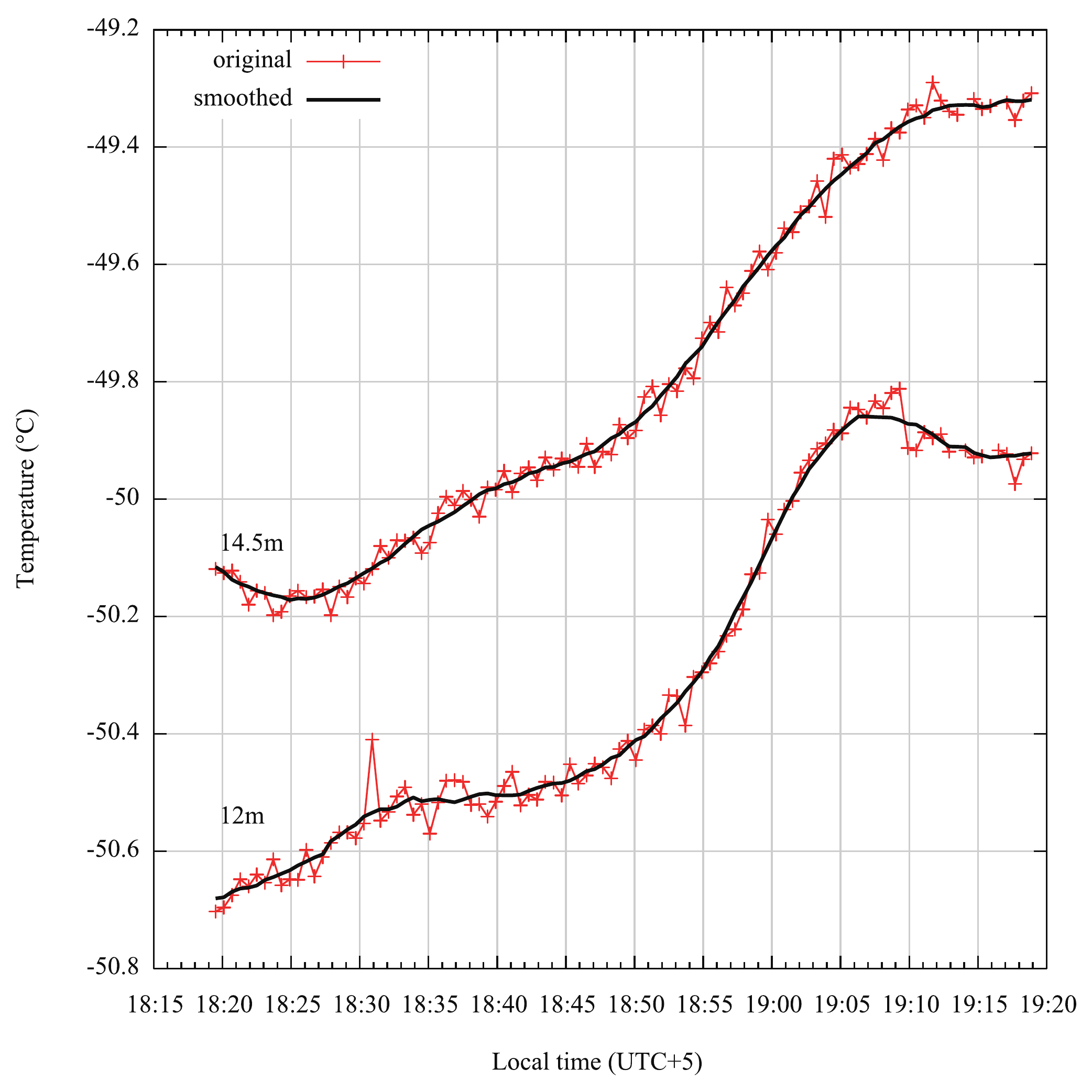}
\caption{An example of temperature data smoothing for the 14.5\metre\ and 12\metre\ sensors for a one hour period on May 15, 2011. Note that the reproducibility of individual measurements on these timescales is much better than the absolute sensor accuracy of $0.32^{\circ}C$.}
\label{fig:fig4}
\end{figure*}

\begin{figure}
\includegraphics[width=0.6\columnwidth]{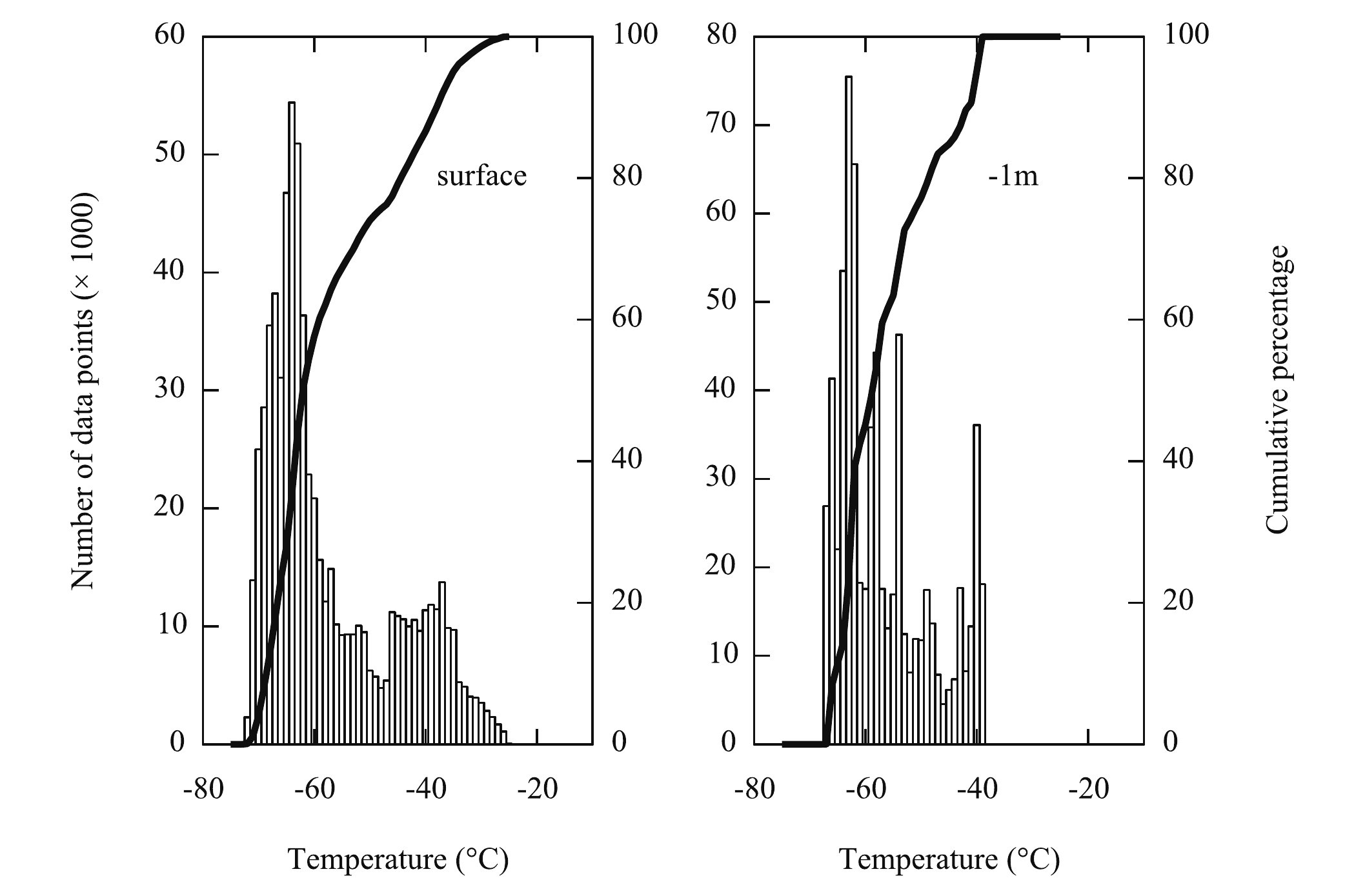}
\caption{Histograms and cumulative distributions (solid lines) of the
  temperature at the surface level and 1\metre\ under the snow surface during 2011.}
\label{fig:fig5}
\end{figure}

\begin{landscape}
\begin{figure}
\includegraphics[width=\columnwidth]{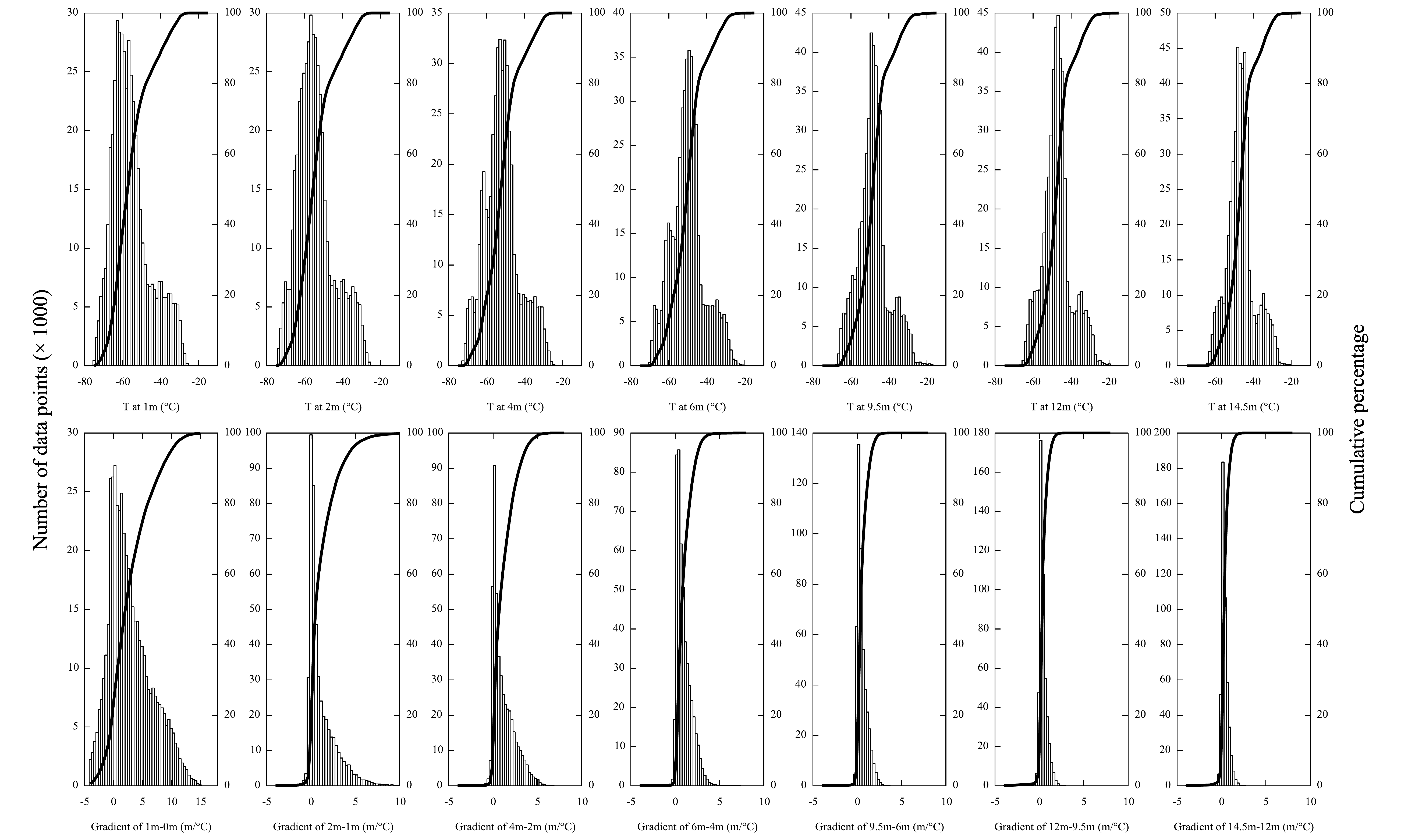}
\caption{Histograms and cumulative distributions (solid lines) of the
  temperatures at 1\metre, 2\metre, 4\metre, 6\metre, 9\metre, 12\metre\ and 14.5\metre\ during 2011, and the
  temperature gradients between every two adjacent levels.}
\label{fig:fig6}
\end{figure}
\end{landscape}

Using data over the entire year, we show the temperature distributions
at snow surface level and at $-1$\metre\ (1 meter below surface) in
Fig.\@ \ref{fig:fig5}.  As expected, the snow temperature at
$-1$\metre\ has a much smaller range than the air temperature.  Due to
snow accumulation, we believe the sensor at the snow surface level was
buried under snow later in the year and could not provide accurate air
temperature measurements.  This can be seen clearly in
Fig.~\ref{fig:fig3} when the surface temperature line (blue) becomes
very smooth in the lower panel.   
We note the temperature distribution of $-1$\metre\ shows multiple
spikes in the right panel of Fig.\@ \ref{fig:fig5} and they are real. 
By investigating the data, we find that the 
temperature at $-1$\metre\ varied little around some values
for considerable long periods, making such isolated spikes.
The temperature distributions at the other heights are shown in Fig.~\ref{fig:fig6} with all data up to
2011 August 3.  They all show a single peak distribution.

\begin{figure*}
\includegraphics[width=\columnwidth]{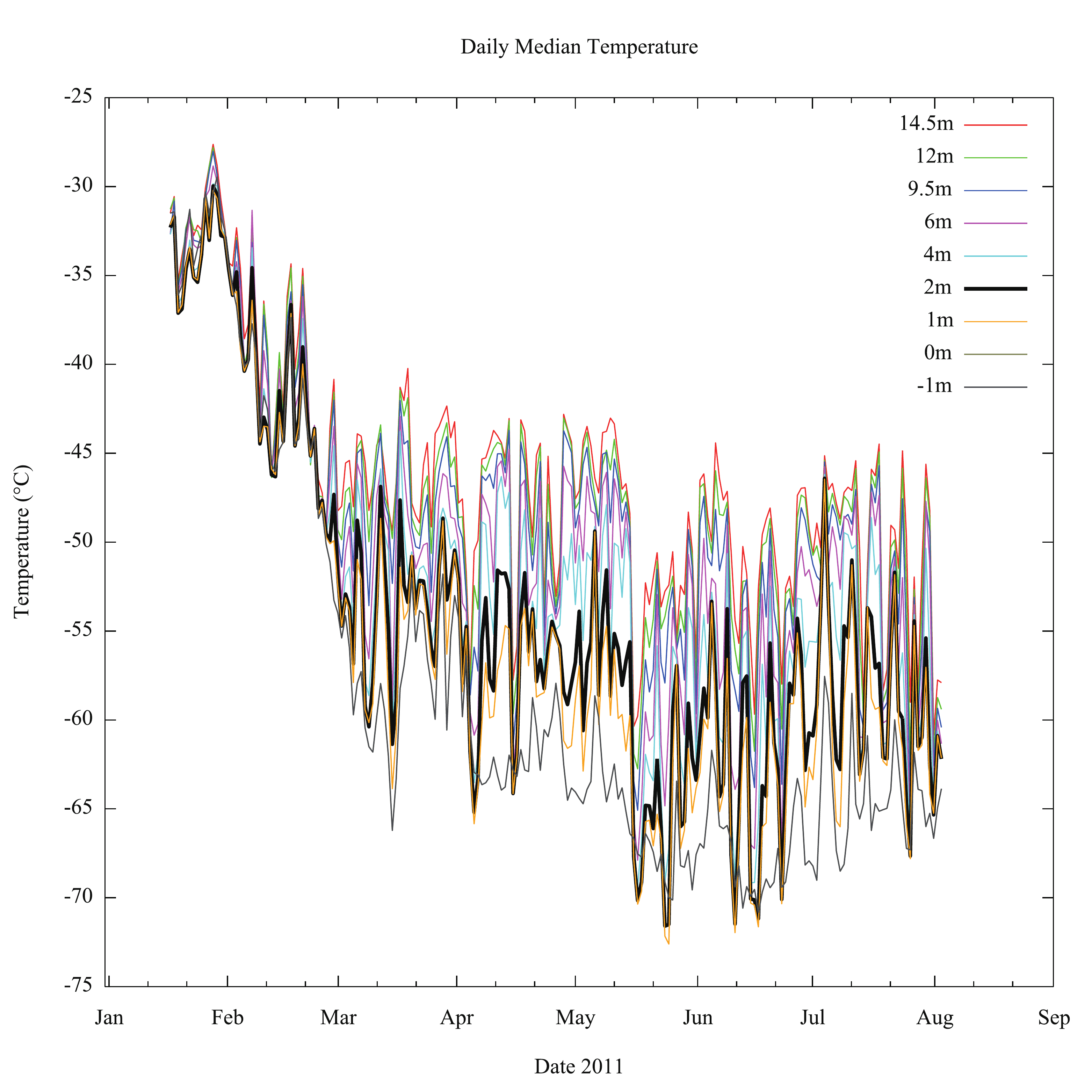}
\caption{Daily median temperatures during 2011. The bold black line is the
temperature at 2\metre.}
\label{fig:fig7}
\end{figure*}

We plot the daily median temperature in Fig.\@ \ref{fig:fig7} to show
the long-term trend which follows the seasonal pattern. The daily
median temperature at 2\metre\ was $\sim-35^{\circ}$C in January, went down
rapidly to  $\sim-60^{\circ}$C in April, and was as low as
$\sim-70^{\circ}$C during the dark winter before it slowly rose again.
Compared with that of South Pole, we find that there is no
significant difference in monthly median temperature between South
Pole and Dome A (see Fig.\@ 2 in Hudson and Brandt 2005). The trend of
the daily median temperature is also similar to that at the South
Pole. 

\begin{figure*}
\includegraphics[width=0.4\columnwidth]{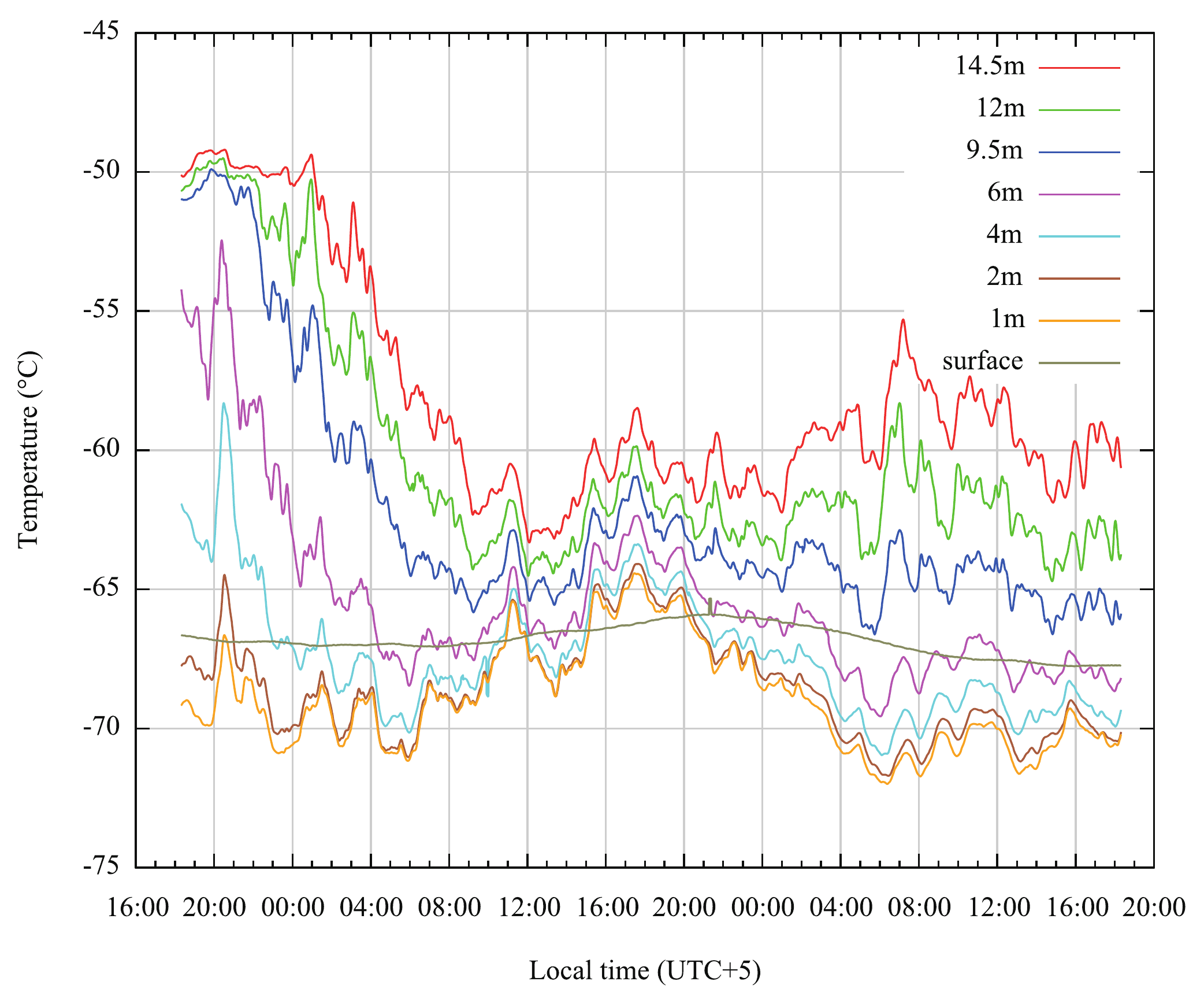}
\caption{Temperatures at different heights between 2011 May 15 and 2011 May 17. 
A positive temperature gradient existed as low as 1\metre\
and lasted for more than two days at all levels. This is indicative of very 
low atmospheric turbulence.}
\label{fig:fig8}
\end{figure*}

Because of the radiatively-cooled snow surface, it is not surprising
that there is strong temperature inversion just above the snow surface
of Dome A.  Fig~\ref{fig:fig8} shows
a good example of the temperature inversion at different heights.  We
define the temperature differences at 1\metre, 2\metre, 4\metre,
6\metre, 9.5\metre, 12.5\metre\ and 14.5\metre\ as the temperature at
the corresponding height minus the temperature at the adjacent lower
height. For example, $\Delta T(2\hbox{\rm m}) = T(2\hbox{\rm m}) -
T(1\hbox{\rm m})$.  
We then calculate the temperature gradients (in $^{\circ}$C/m) at
different heights and their distributions are 
are shown in Fig.\@ \ref{fig:fig6}.  The distributions are relatively
narrow except for the one involving $\Delta T(1m)$, since the surface
temperature that did not always measure the air temperature as
explained in \S3.1.  For all other heights, the distributions increase
dramatically from negative values to a positive peak 
and go down relatively slowly with a tail. 
The temperature gradients are smaller at higher elevations.
The existence of the positive temperature gradients indicates strong
temperature inversion at all layers.  


To compare the temperatures at two levels spanning more than
10\metre\ in height, Hudson and Brandt (2005) found at South Pole that
for 99.9\% of the time there was a positive temperature difference
between 13\metre\ and 2\metre.  We found the same for our data at Dome
A between 12\metre\ and 2\metre.  However, the median temperature
difference at Dome A was $6.9^{\circ}$C, which is much larger than the
difference of $1.1^{\circ}$C at South Pole. This shows that the
strength of the temperature inversion at Dome A is much greater than
that at South Pole.

\begin{figure*}
\includegraphics[width=0.4\columnwidth]{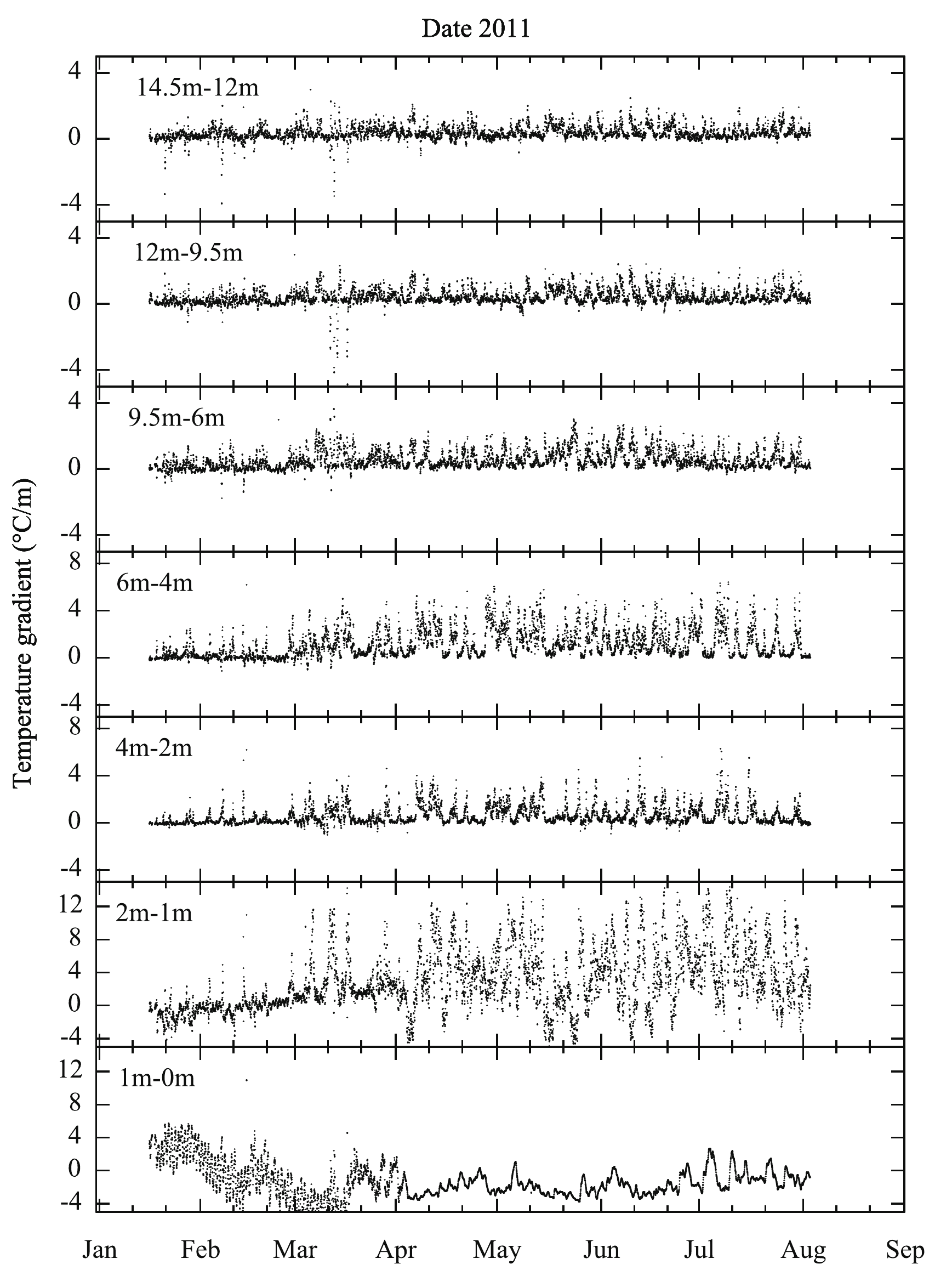}
\caption{The temperature gradient between 2011 January and August at all levels. The temperature gradients were calculated from the smoothed data (see \S2). And the data points were down-sampled by 100 for clarity.}
\label{fig:fig9}
\end{figure*}

We plot the temperature gradients at all levels between 2011 January and August
in  Fig.\@  \ref{fig:fig9}.  This figure shows
that most of the large negative temperature gradients happened during
January and February while most of the large positive gradients
happened in the wintertime.  Higher layers tend to have smaller
temperature gradients than lower layers as also seen in
Fig.~\ref{fig:fig6}.

\begin{landscape}
\begin{figure*}
\includegraphics[width=\columnwidth]{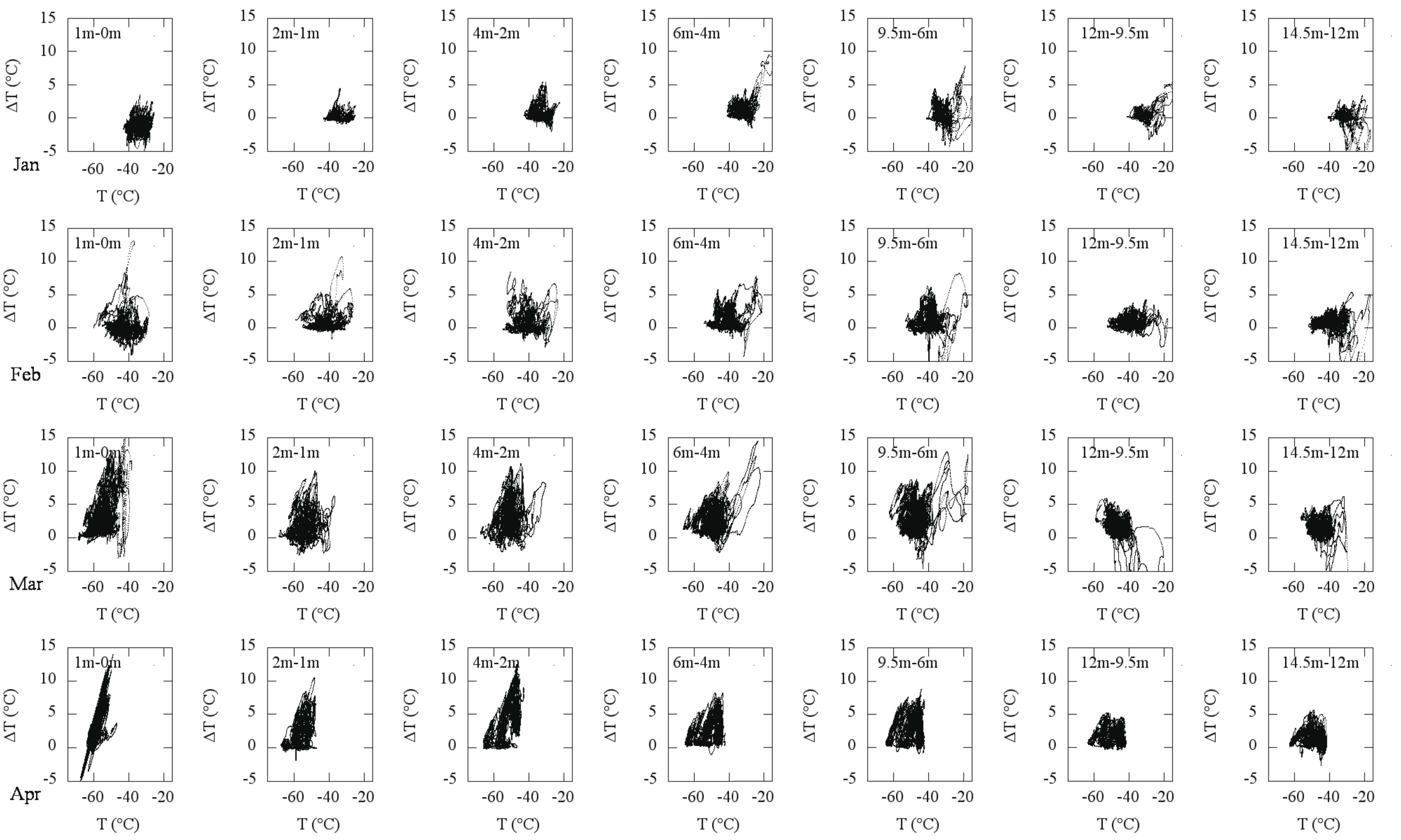}
\caption{Temperature difference vs. temperature in 2011 January, February, March and April.}
\label{fig:fig10}
\end{figure*}
\end{landscape}

\begin{landscape}
\begin{figure*}
\includegraphics[width=\columnwidth]{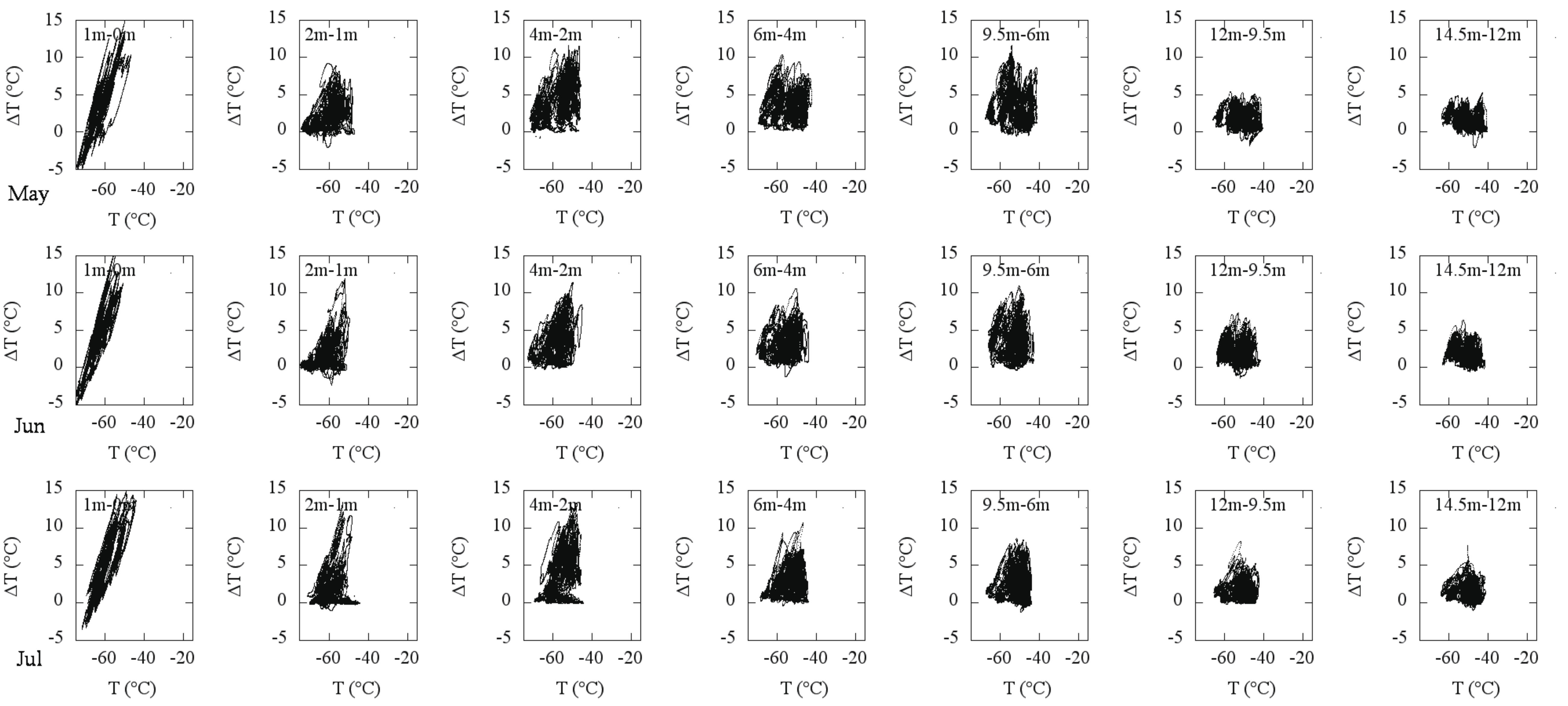}
\caption{Temperature difference vs. temperature in 2011 May, June and July.}
\label{fig:fig11}
\end{figure*}
\end{landscape}

We also investigated whether the temperature inversion depends on the
temperature or season.  Figs.~\ref{fig:fig10} and \ref{fig:fig11} show
the temperature difference and temperature for different months during
2011.  It is clearly seen that most of the largest positive temperature
differences (e.g. $>5^{\circ}$C) occurred when the temperature was between
$-45^{\circ}$C and $-65^{\circ}$C.  The strong correlations between
$\Delta T$ and $T$ at 1\metre\ for later months are not real and are again
caused by the snow buried sensor at 0\metre\ measuring the snow temperature,
which did not vary as much as the air temperature.  The other
plausible correlations for 2\metre\ and 4\metre\ after March could be real,
suggesting that at lower layers a larger temperature inversion occurred
at higher temperatures.
There is a strong trend that after March, at 4\metre, 6\metre, and
9\metre, the temperature differences are almost all positive,
indicating long durations of temperature inversion.

\begin{figure*}
\includegraphics[width=0.9\columnwidth]{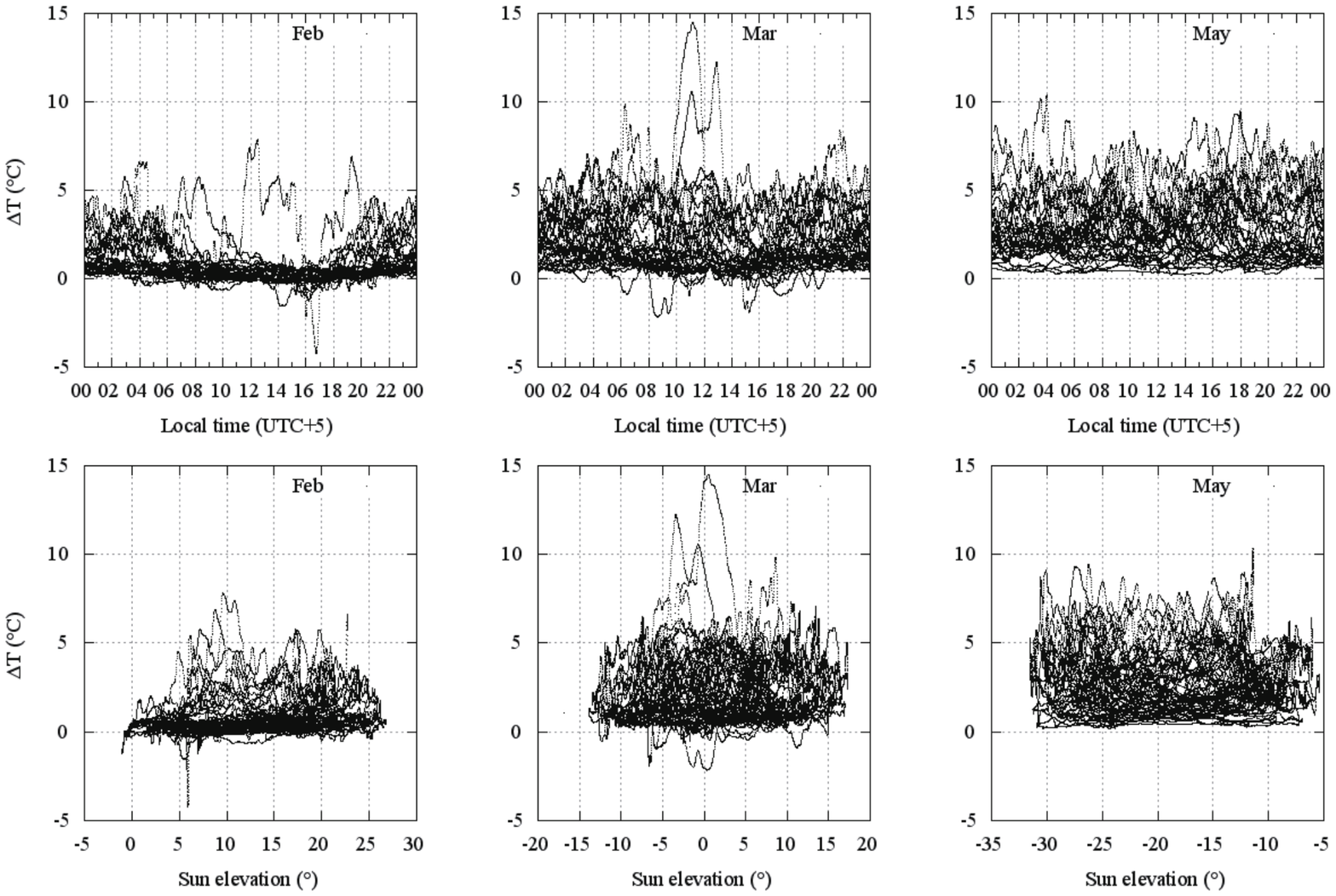}
\caption{Temperature difference between 6\metre\ and 4\metre\ folded into a 
day during February, March, and May (top) and the same temperature
difference vs. solar elevation (bottom).}
\label{fig:fig12}
\end{figure*}

Although the daily temperature obviously varies with the elevation of the Sun
in the summer season as seen in the top panel of Fig.\@ \ref{fig:fig3},
we found that the temperature difference in a day has little
relationship with the Sun.  In Fig.\@ \ref{fig:fig12} we fold the
temperature difference at 6\metre\ into one day for February, March and
May, and also plot the temperature difference against the solar
elevation.  In February of 2011, the temperature difference at 6\metre\ seems to
be larger at night and seems to be related to the Sun, but in March
when the Sun still rises and sets at Dome A, we do not find a preferable
time for larger temperature differences and there is no evidence of
correlation between $\Delta T$ and solar elevation.

\begin{figure*}
\includegraphics[width=0.5\columnwidth]{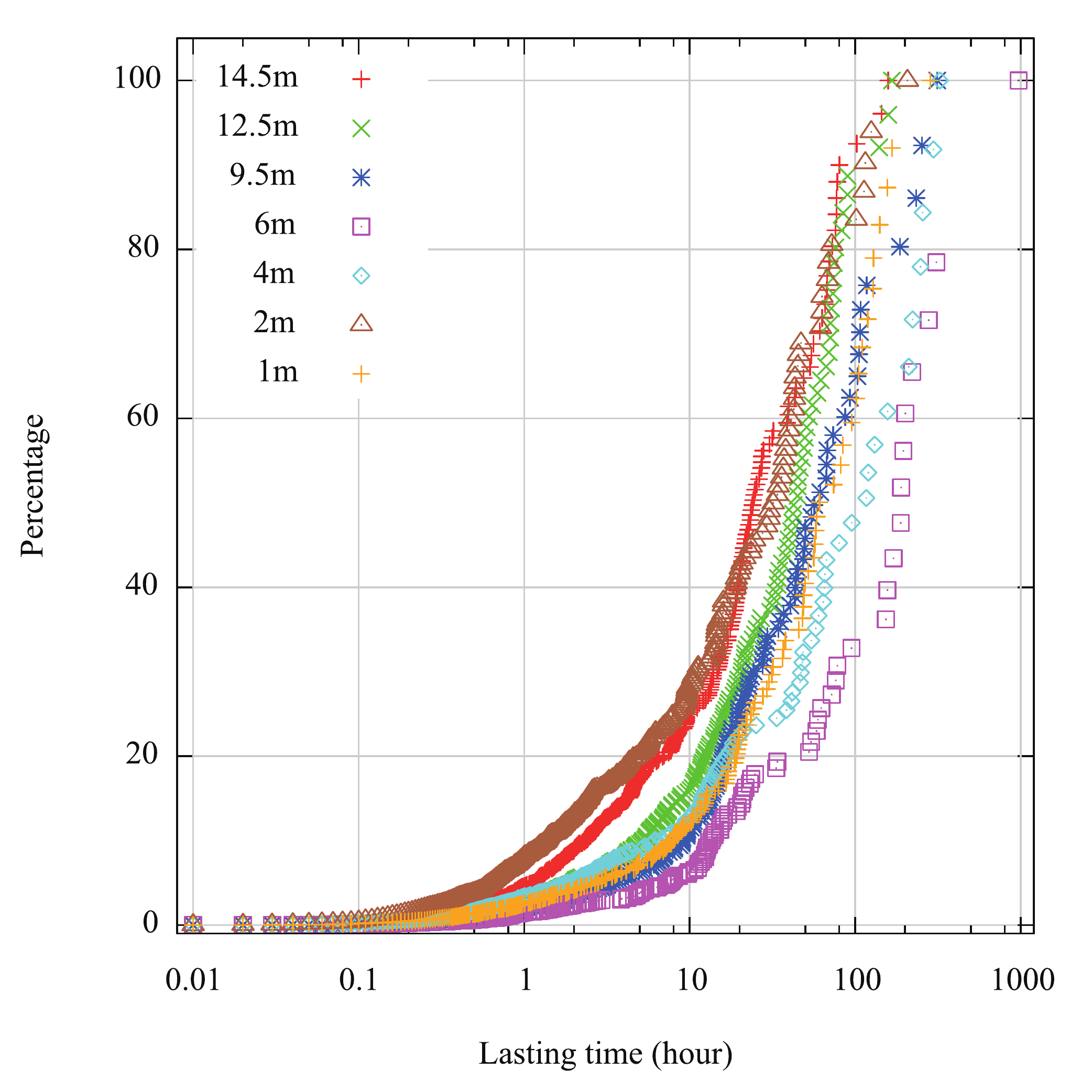}
\caption{Cumulative distributions of the duration of the temperature
  inversion (defined as a positive temperature difference larger than
  $0.14^{\circ}$C) for all heights.}
\label{fig:fig13}
\end{figure*}

As an indicator of the stability of the atmosphere, we are mostly
concerned with how long a temperature inversion lasts. We show a
clear example of two days' data in mid May, 2011 in Fig.\@
\ref{fig:fig8} where a  temperature inversion exists at all heights.
We evaluated the temperature inversion (i.e., temperature difference) at each
height with respect to the temperature sensor immediately below.
To evaluate the duration of a temperature inversion at all heights, we need to
define a real temperature difference quantitatively.  While the temperature
sensors have a specified absolute accuracy of $0.32^{\circ}$C, our data
show that the variation in consecutive measurements is much smaller
than this (see Fig.\@ \ref{fig:fig4}).  We have subtracted the
smoothed data from the original data and calculated the standard
deviation as $\sigma=0.033^{\circ}$C.  Therefore, we set a $3\sigma$
temperature difference threshold of $0.14^{\circ}$C, above which we
regard it as a real temperature inversion rather than random
measurement error. We choose all the temperature difference data
points larger than $0.14^{\circ}$C, and counted how long the inversion lasted.
The results are shown in Fig.\@ \ref{fig:fig13} and
Table~\ref{tab:duration}.  
A temperature inversion existed more than 70\% of the time at all
heights, and up to 95\% at a height of 6\metre.  We note that when a temperature
inversion occurred, it lasted at least 25 hours for more than
$\sim$40\% of the time at all heights and up to 78\% at 6\metre.  For
less than 10\% of the time, the temperature inversion lasted under 1~hour.

We also note that the median duration of the temperature inversion
increased from 2\metre\ to 6\metre\ and then decreased with height.  There
are two possible reasons for this trend.  One is that the air
turbulence is stronger at the lower elevation, the other is that
sudden strong wind can destroy the temperature inversion at higher
elevations (see the next subsection for wind speed statistics).

\subsection{Wind speed and direction}

\begin{figure*}
\includegraphics[width=\columnwidth]{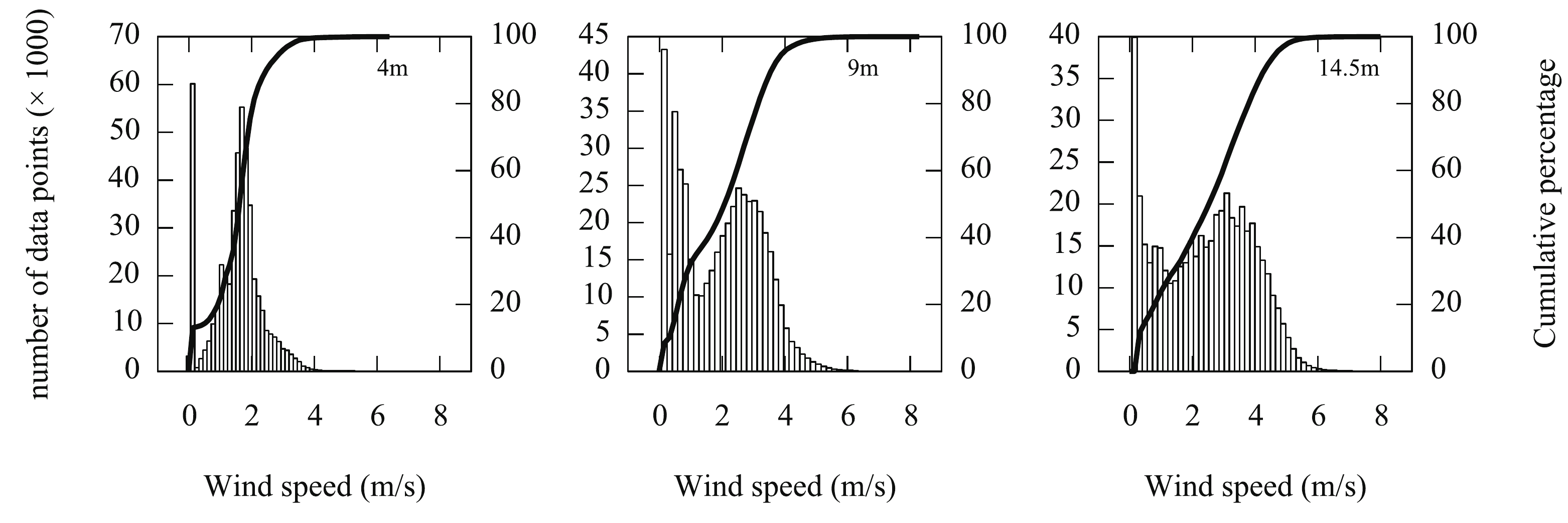}
\caption{Wind speed histograms and cumulative distributions (solid lines) at 4\metre, 9\metre\ and 14.5\metre\ during 2011.}
\label{fig:fig14}
\end{figure*}

\begin{figure*}
\includegraphics[width=\columnwidth]{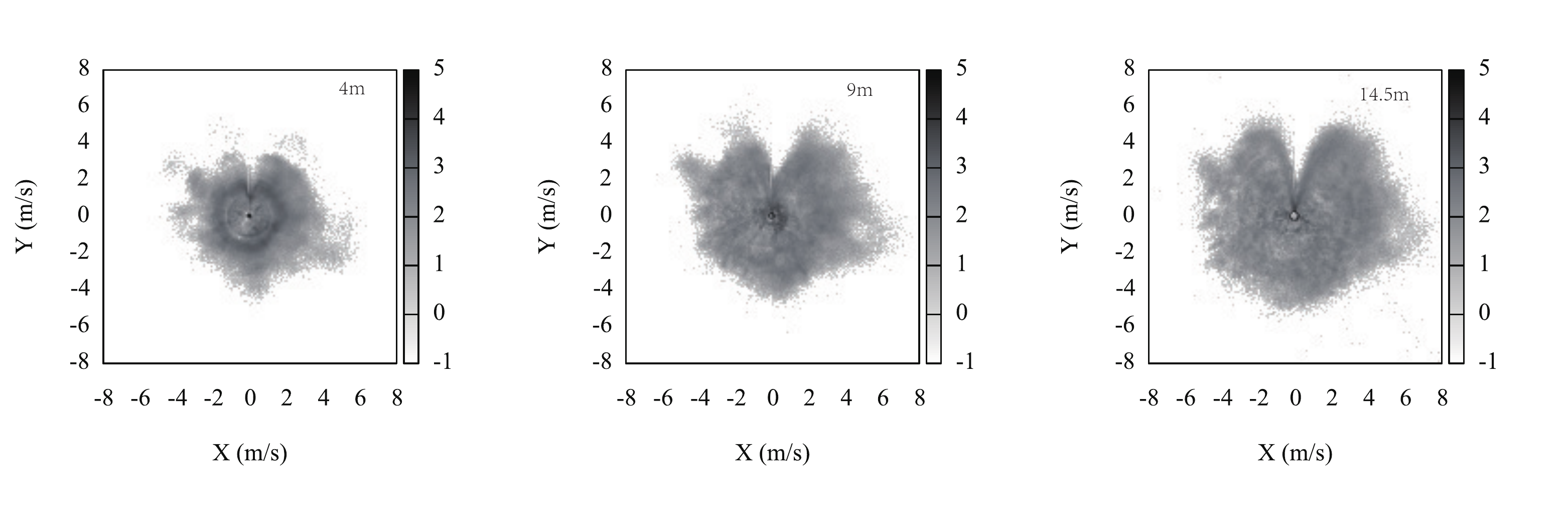}
\caption{The wind rose density at three heights during 2011. North is up.  The gaps are caused by the weather mast which blocked the wind from the north direction. The grey scale is the logarithm of the number of data points per pixel. 
The obvious darker ring in the 4m plot corresponds to the dominant
wind speed around about 1.7\mps\ (see Fig.~\ref{fig:fig14}) 
of random directions.  Similar
patterns also exist in the other two plots, but not as obvious.}
\label{fig:fig15}
\end{figure*}

The distributions and cumulative statistics for the wind speed at three
heights are shown in Fig.\@ \ref{fig:fig14}.  The peak at 0\mps,
accounting for about 10\% of the time, simply means there was no
wind, or the wind speed was too low to be detected by the sensors,
or the sensors got stuck (see below).
At 4\metre, the wind speed was lower
than 1.5\mps\ for half of the time, and seldom reached 4\mps.  The
wind speed increases with height above the snow.  At 9\metre\ and
14.5\metre, the median wind speeds are 2.0\mps\ and 2.4\mps,
respectively, and could often reach 4\mps.  This can also be clearly
seen in the wind rose plots (Fig.~\ref{fig:fig15}).  We notice that
there is no dominant wind direction for all three heights.  The
obvious gaps in the wind roses are caused by the mast of KL-AWS which
blocked wind in that direction.

\begin{figure*}
\includegraphics[width=0.5\columnwidth]{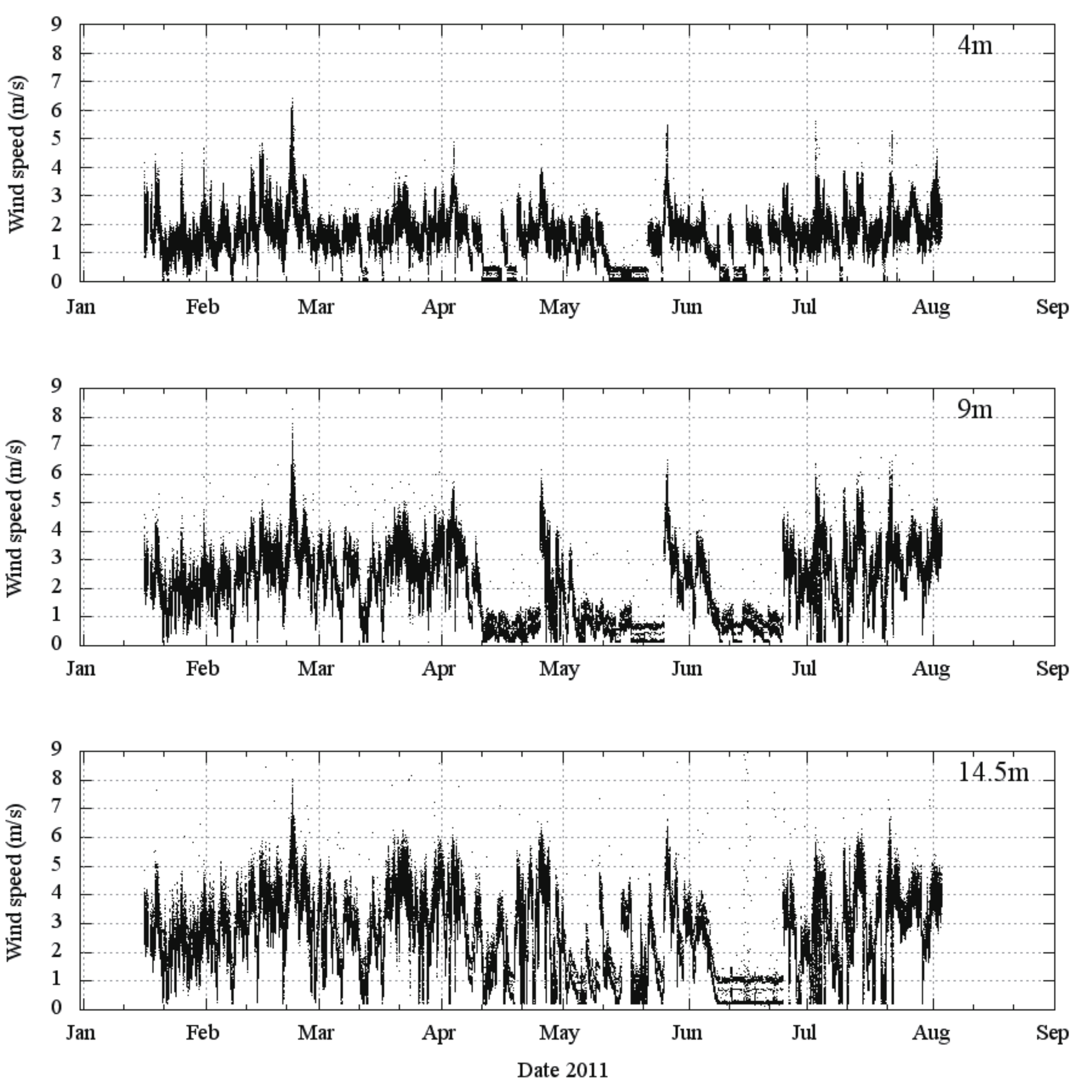}
\caption{Wind speed during 2011 at three heights. Unsmoothed raw data are shown in this figure.}
\label{fig:fig16}
\end{figure*}

\begin{figure*}
\includegraphics[width=0.5\columnwidth]{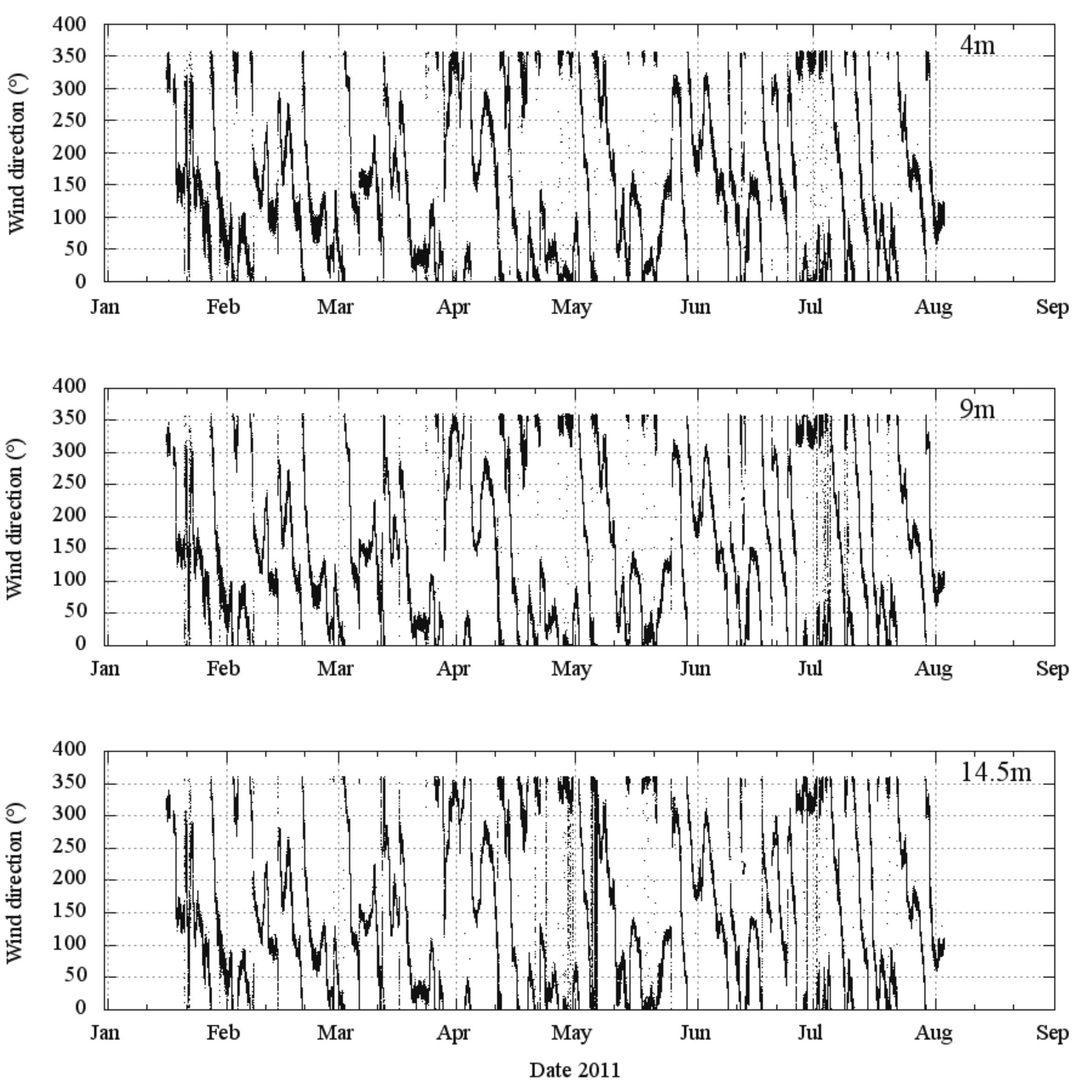}
\caption{Wind direction during 2011 at three heights. Unsmoothed raw data are shown in this figure.}
\label{fig:fig17}
\end{figure*}

We have shown the wind speed and direction between January and August
in Fig.~\ref{fig:fig16} and Fig.~\ref{fig:fig17}, and noticed that
when there is virtually no wind, it was mostly during the polar night
when the Sun never rose.  However, for some extended periods during
April to June, the wind speeds were flat in the plots and were 0\mps\ 
or very low.  This
makes us to worry that it is possible that the propellers of the
anemometers got stuck sometimes at low temperatures.
Fig.~\ref{fig:fig17} shows again that the wind directions are random
and they are synchronized very well at the three heights.

We have compared the average wind speeds at Dome A, Dome C, and the
South Pole in Table \ref{tab:tab2} and find that the horizontal wind
speed at Dome A at 4\metre\ was only 1.5\mps, much smaller than the
ground wind speed of 2.9\mps\ at Dome C and 5.5\mps\ at the South
Pole. 

\begin{figure*}
\includegraphics[width=\columnwidth]{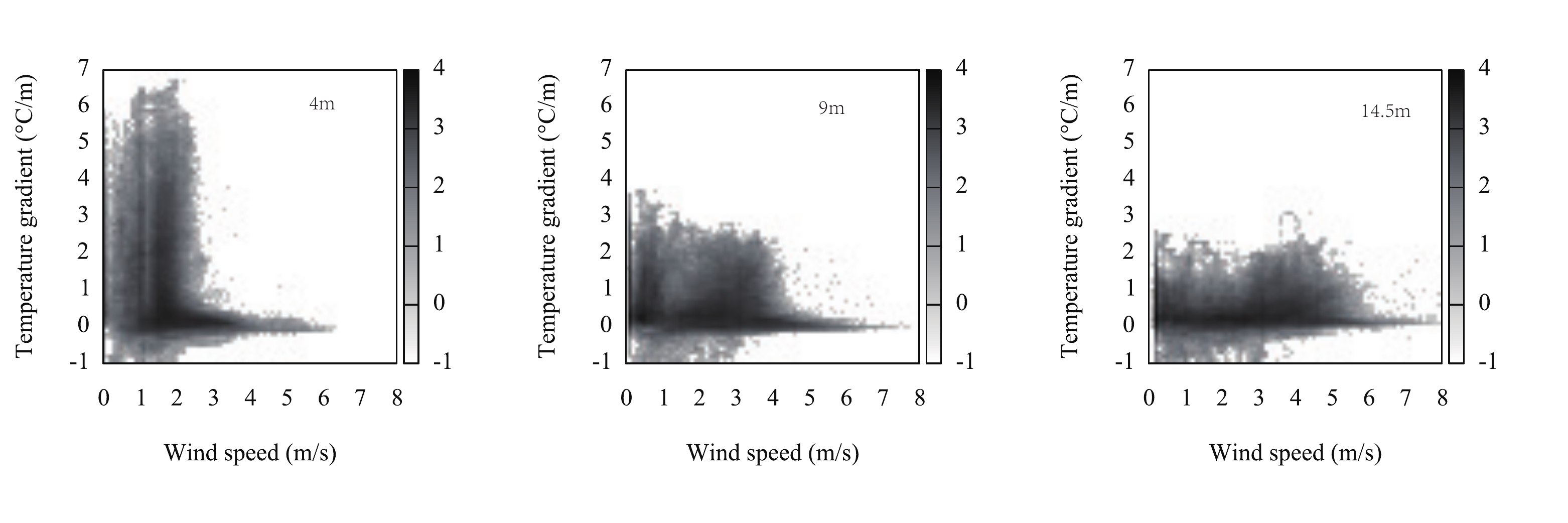}
\caption{Wind speed against temperature gradient at the corresponding height. The grey scale is the logarithm of the number of data points per pixel. 
The vertical structures in the plots correspond to the real
distribution of the data points in Fig.~\ref{fig:fig14}.}
\label{fig:fig18}
\end{figure*}

In order to investigate whether the temperature gradient is related to
wind speed, we plot the wind speeds at 4\metre, 9\metre\ and
14.5\metre\ against temperature gradients at 4\metre, 9.5\metre\ and
14.5\metre\ respectively in Fig.\@ \ref{fig:fig18}.  Stronger
temperature inversion always occurred when the wind speed was lower.
When the wind speed is higher than 2.5\mps\ at 4\metre, the temperature
gradient was never larger than $0.7^{\circ}$C/m.  We also see the same
trends at 9\metre\ and 14.5\metre, but with different specific wind
speeds of 4.1\mps\ and 5.3\mps, respectively. It therefore seems that
at the higher elevations, the positive temperature gradients are
smaller, but more resistant to wind.

\subsection{Air pressure}

\begin{figure*}
\includegraphics[width=\columnwidth]{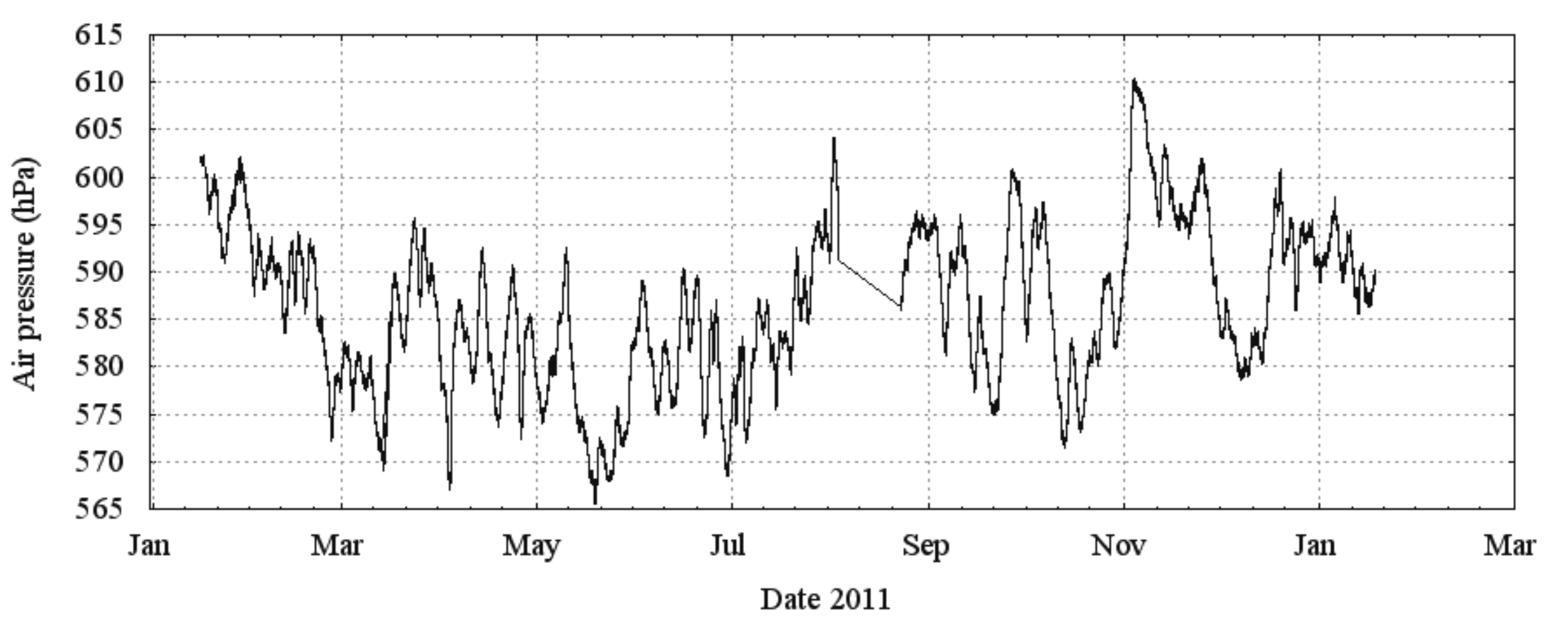}
\caption{Air pressure during 2011. The raw data for the entire year are shown in this figure. Note that the barometric data should not be influenced by the collapse of the upper part of the mast in September.}
\label{fig:fig19}
\end{figure*}

\begin{figure*}
\includegraphics[width=0.5\columnwidth]{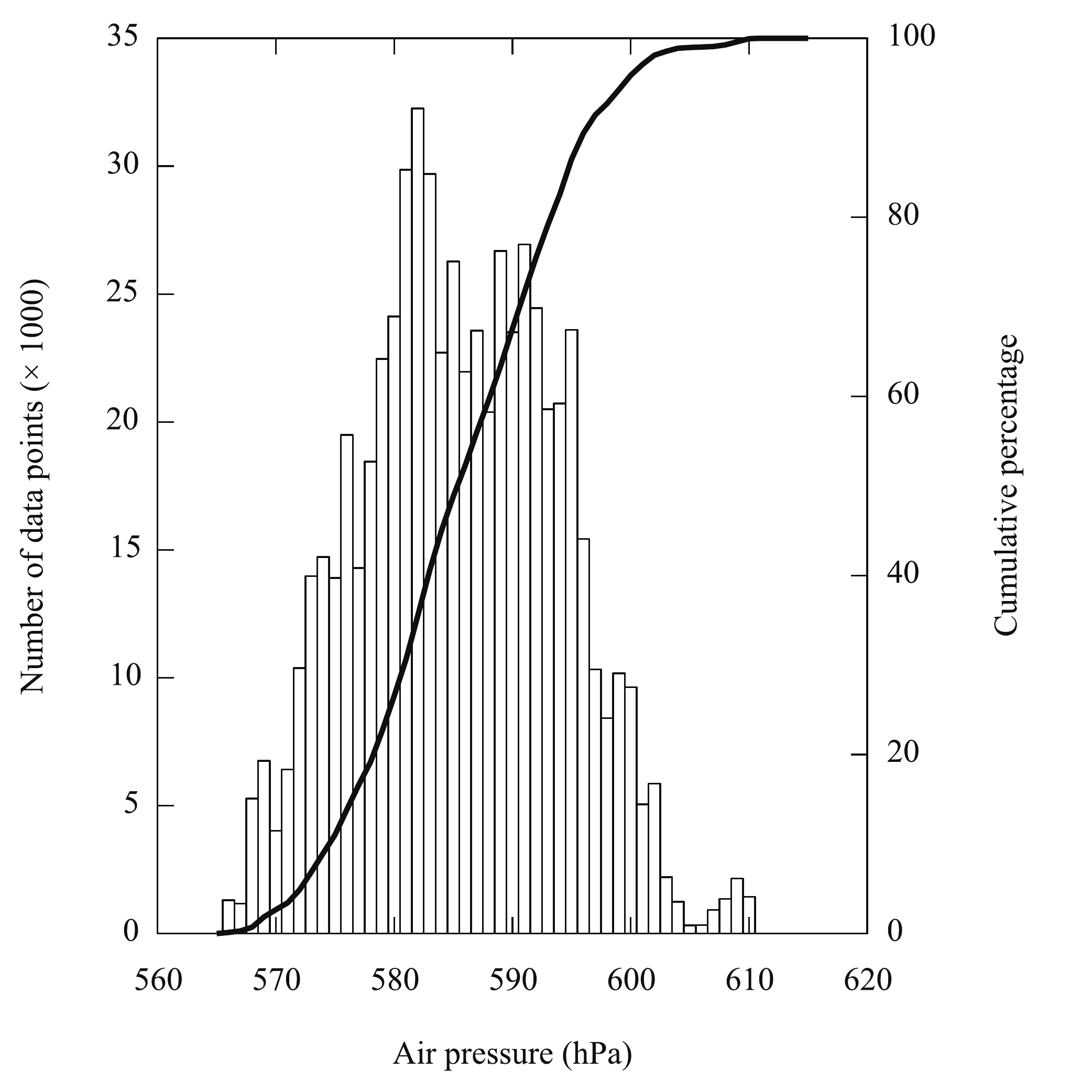}
\caption{Air pressure histogram and cumulative distribution (solid line) during 2011.}
\label{fig:fig20}
\end{figure*}

\begin{figure*}
\includegraphics[width=\columnwidth]{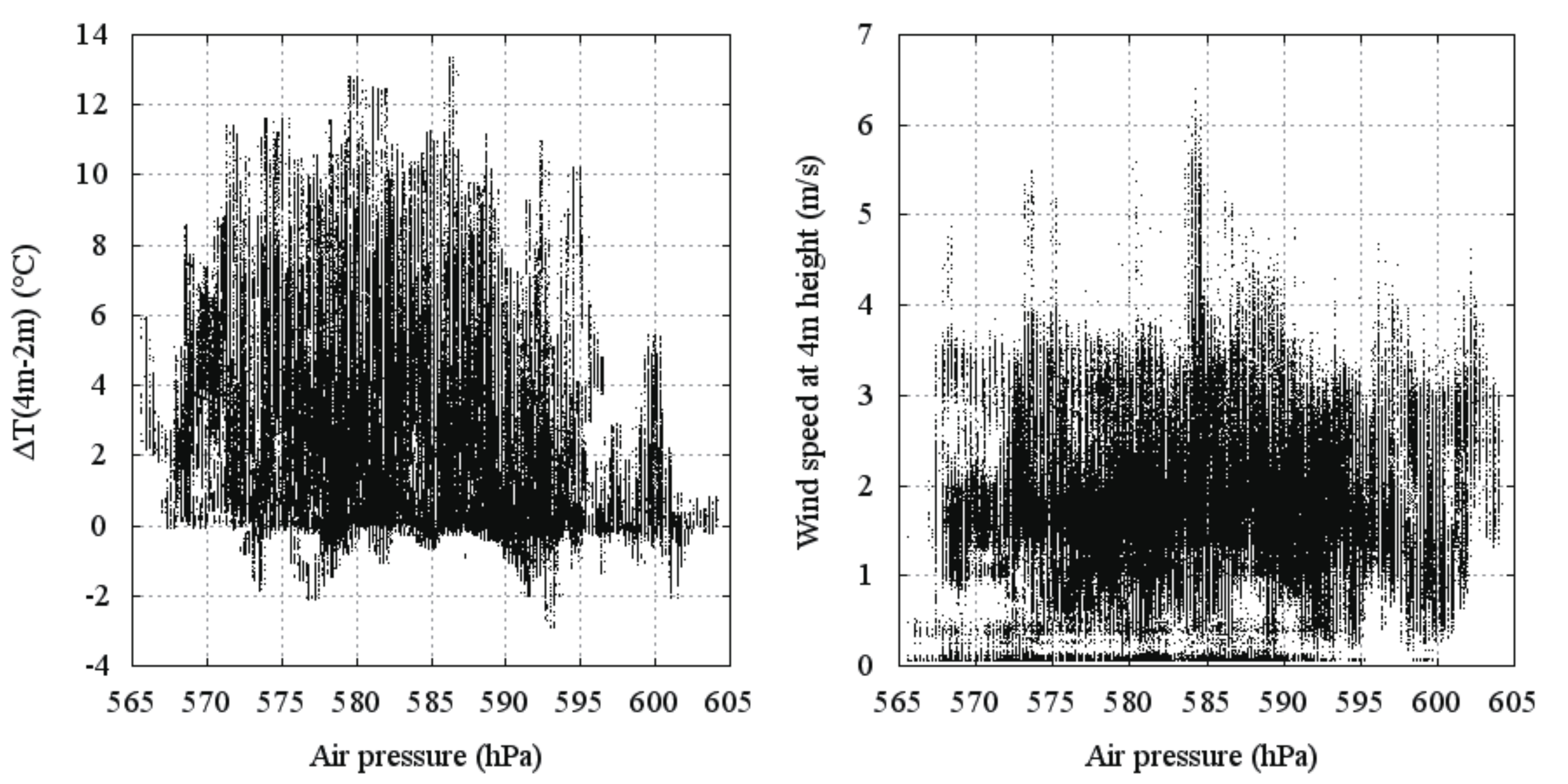}
\caption{Temperature difference between the 2\metre\ and 4\metre\ heights (left panel) and
wind speed at 4\metre\ (right panel) versus air pressure during 2011. }
\label{fig:fig21}
\end{figure*}

The barometer at the height of 1\metre\  worked correctly for the entire year.
Fig.~\ref{fig:fig19} and \ref{fig:fig20} shows the data and the
distribution. The average air pressure at Dome A was
586\,hPa with a standard deviation of 8\,hPa. 
We do not find any clear relationships between the air pressure and the
temperature difference or wind speed at 4\metre\ (Fig.~\ref{fig:fig21}).

\section{Conclusion and discussion}

We have analyzed the weather data collected from KL-AWS at Dome A
during 2011.  The average temperature between January and July was
$-54^{\circ}$C at 2\metre\ and $-46^{\circ}$C at 14.5\metre.  We have
found that there is usually a strong temperature inversion at all the
heights above the ground.  The temperature inversion, defined here as
a positive temperature difference larger than $0.14^{\circ}$C, existed
more than 70\% of the time at all heights and up to 95\% at 6\metre.
Such inversions lasted longer than 25 hours for more than about 40\%
of the time at all heights, and almost 80\% of the time at 6\metre.
Most largest negative temperature differences (no
inversion) occurred during January and February and most strong
temperature inversion occurred when the temperature fell between
$-65^{\circ}$C and $-45^{\circ}$C.

The average wind speed at 4\metre\ is 1.5\mps, which may be the lowest
reported from any site on earth. The wind speed at higher elevation is
slightly stronger.  We also notice that for $\sim10\%$ of the time
there is almost no wind or the wind speed was too low to be detected
by the sensors, this usually happens during the dark polar night.
However, there also seems to be evidence that the wind speed sensors might
have been stuck sometimes at low temperatures.
The wind on the Antarctic continent is dominated by katabatic winds, therefore the wind speed
should be very small and the wind direction should be random above the
highest point (Dome A).  Hudson and Brandt (2005) found that the
strongest temperature inversion often happened with a 3--5\mps\ wind at
the South Pole, but we find that strong temperature inversion happens
when the wind speed is low, such as less than 2.5\mps\ for the height of
4\metre; when the wind speed is higher than 2.5\mps\ the temperature gradient
decreases sharply.

\begin{figure*}
\includegraphics[width=0.5\columnwidth]{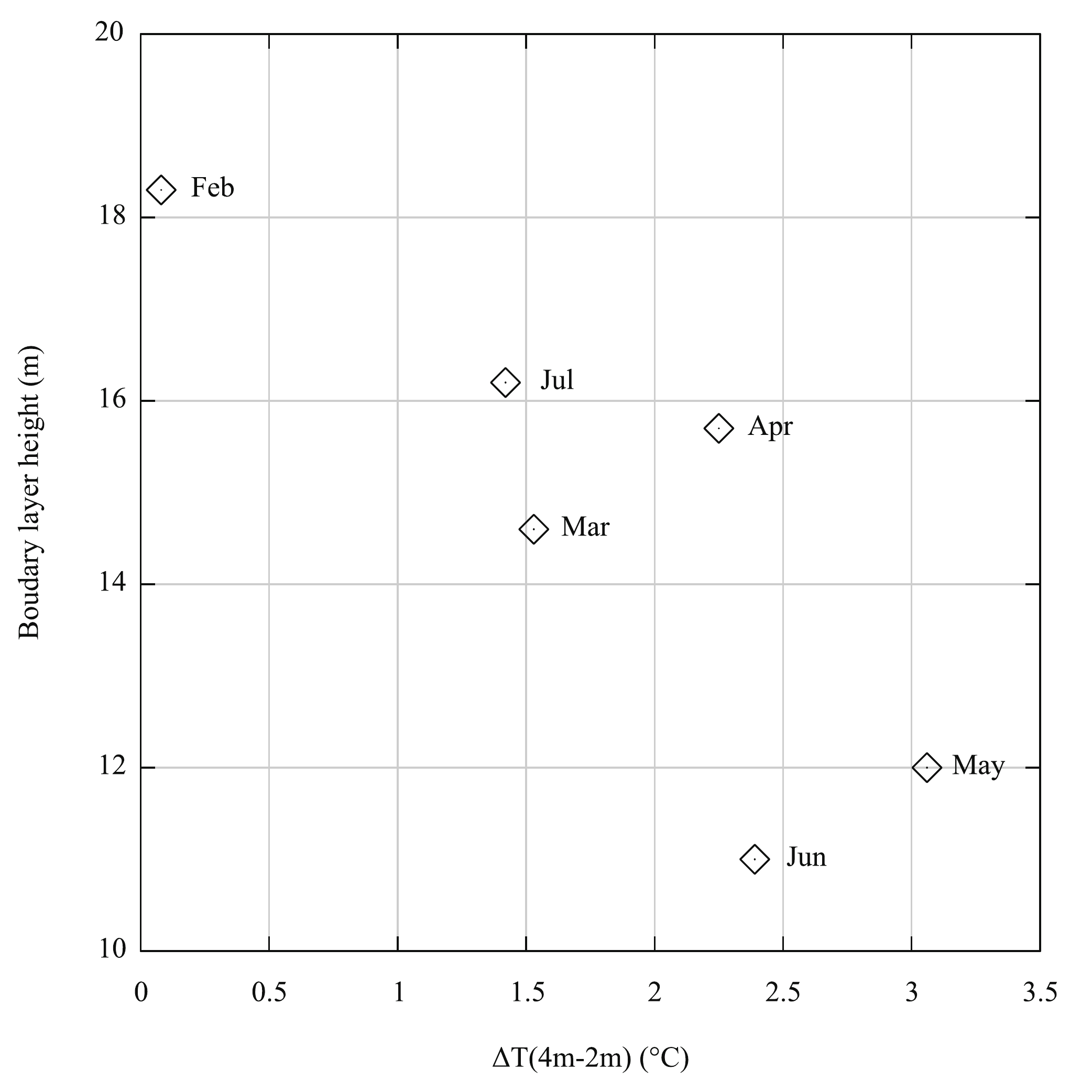}
\caption{Monthly median boundary layer height during 2009 (Bonner
  \etal\ 2010) versus the monthly median temperature difference
  between heights of 2\metre\ and 4\metre\ during 2011.}
\label{fig:fig22}
\end{figure*}

It has been known for a long time that the surface of the Antarctic
plateau is prone to temperature inversions due to radiative cooling of
ice, and hence a lower air temperature near the surface.
As Basu and Port\'{e}-Agel (2006) pointed out, whenever a temperature
inversion exists, turbulence is generated by shear and destroyed by
negative buoyancy and viscosity. Competition between shear and
buoyancy weakens turbulence and results in a shallower boundary layer.
We list the monthly median boundary layer heights of Dome A in 2009
(Bonner \etal\ 2010) and the monthly median temperature difference at
4\metre\ during 2011 in Table \ref{tab:tab3} and compared them in
Fig.~\ref{fig:fig22}.
If average conditions were similar between 2009 and 2011, this
indicates, as expected, that stronger temperature inversions favor a
thinner boundary layer, with the lowest in May and June.

The strong temperature inversion and low wind speed at Dome A both
produce a very stable atmosphere and a stable boundary layer with
little convection, especially in winter.  When the temperature
inversion happens as low as 2\metre, we expect the boundary layer can
be much lower too.  Above the shallow boundary layer, we can reach the
free atmosphere where the seeing will be good for astronomical
observations. In other words, there will be periods when even
telescopes that are very close to the ground will experience
free-atmosphere seeing approaching 0.25\arcseconds.

We have demonstrated that the weather data collected by KL-AWS at
different heights are very useful to characterize the stability of the
near-ground atmosphere at Dome A and can provide insights for evaluating
the site for astronomical observations.  We plan to continue the
experiment with a new weather tower and monitor the site for at least
2--3 years so that we can have better statistics and understanding
of the site.

The data from KL-AWS are available in raw form at
\url{http://aag.bao.ac.cn/weather/downloads/} and as plots at
\url{http://aag.bao.ac.cn/weather/plot\_en.php}

\acknowledgments 

The authors wish to thank all members of the 27th and 28th Chinese
Antarctic Research Expedition teams for their effort in setting up the
KL-AWS instruments and servicing the PLATO observatory.  This research
is supported by Chinese Polar Environment Comprehensive Investigation
\& Assessment Programmes under grant No.~CHINARE2012-02-03, National
Basic Research Program of China (973 Program 2013CB834900), and the
National Natural Science Foundation of China under grant No.~11273019
and 11203039.
This project is also supported by the Commonwealth of Australia under
the Australia-China Science and Research Fund, and by the Australian
Research Council and the Australian Antarctic Division.

\begin{table}
\caption{Cumulative percentage of the duration of temperature inversion}
\label{tab:duration}
\begin{center}
\begin{tabular}{r|ccc}
\hline

Height  &  $>25$ hours & $>10$ hours  & Total\\
\hline
 1\metre\     &  55  &  66  &  75 \\ 
 2\metre\     &  39  &  51  &  71 \\ 
 4\metre\     &  64  &  73  &  84 \\ 
 6\metre\     &  78  &  89  &  95 \\ 
 9.5\metre\   &  59  &  76  &  85 \\ 
 12\metre\    &  57  &  73  &  86 \\ 
 14.5\metre\  &  41  &  64  &  85 \\ 
\hline
\end{tabular}
\end{center}
\tablecomments{The temperature inversion is evaluated at each height
with respect to the temperature sensor immediately below (\S\ref{sec:t}).}
\end{table}

\begin{table}
\caption{Average wind speed at Dome A, Dome C and the South Pole.}
\label{tab:tab2}
\begin{center}
\begin{tabular}{ccl}
\hline
Site & Wind speed (\mps)  & Reference \\
\hline
Dome A (4\metre) & $ 1.5 $ & This paper \\
Dome C (3.3\metre) & $2.9$ & Aristidi \etal\ (2005) \\
South Pole (2\metre) & $5.5$ & Mefford (2004) \\
 \hline
\end{tabular}
\end{center}
\end{table}
 
\begin{table}
\caption{Dome A monthly median boundary layer height of 2009 (Bonner
  \etal\ 2010) and monthly median temperature difference of 2011
  between 4\metre\ and 2\metre.}
\label{tab:tab3}
\begin{center}
\begin{tabular}{ccc}
\hline
Month &$\Delta T$ (4\metre$-$2\metre)&Boundary layer height\\
&$^{\circ}$C&m\\
\hline
January & $ 0.18 $ & \nodata \\
February & $0.08$ & $18.3$ \\
March & $1.53$ & $14.6$ \\
April & $2.25$ & $15.7$ \\
May & $3.06$ & $12.0$ \\
June & $2.39$ & $11.0$ \\
July & $1.42$ & $16.2$ \\
August & \nodata & $10.2$ \\
 \hline
\end{tabular}
\end{center}
\end{table}

\clearpage

\onecolumn

\clearpage

\end{document}